\documentclass{article}
\usepackage{hiph-art}
\volnumber{} \issuenumber{} \edyear{}                             
\frompage{} \topage{}                                              
\recrevdate{}                                              
\def\rightmark{}
\def\leftmark{U. Reinosa}
 
\def\sumint{\hbox{$\sum$}  \!\!\!\!\!\!\!\int}

\title{Non-perturbative renormalization of $\Phi$-derivable approximations in theories with multiple fields} 
\authors{ 
{U. Reinosa $^1$%
\index{Reinosa, U.} 
}\\[2.812mm]
{\normalsize
\hspace*{-8pt}$^1$ Institut f\"ur Theoretische Physik, Universit\"at Heidelberg,\\ 
Philosophenweg 16, 69120 Heidelberg, Germany\\[0.2ex] 
}}
 
\abstract{We provide a renormalization procedure for $\Phi$-derivable approximations in theories coupling different types of fields. We illustrate our approach on a scalar $\varphi^4$ theory coupled to fermions via a Yukawa-like interaction. The non-perturbative renormalization amounts to fixing the scalar coupling via a set of nested Bethe-Salpeter equations coupling fermions to scalars.}
\keyword{2PI effective action, renormalization}

\PACS{11.10.Wx,11.15.Tk,11.10.Gh}
 
\makeindex
\begin{document}
 
\maketitle

\section{Introduction}\label{intro}
The two-particle-irreducible (2PI) effective action \cite{Luttinger:1960ua} has recently regained interest in many fields of research such as the QCD phase diagram \cite{Blaizot:1999ip} or the dynamics of quantum fields out-of-equilibrium \cite{Aarts:2002dj}. There are however two major difficulties related to how fundamental properties such as UV finiteness or gauge invariance manifest themselves within a given truncation of the 2PI effective action. Recent progresses have been made concerning renormalization in the case of scalar theories \cite{vanHees:2001ik}. Aside from the question of gauge invariance, an important step towards the understanding of renormalization in the case of gauge theories consists in considering a theory coupling multiple fields.

As an illustration, we consider here a massless fermionic field $\psi$ at finite temperature, coupled to a massless self-interacting scalar field $\varphi$ ($\lambda\varphi^4$ theory) via a Yukawa-like interaction $g\bar{\psi}\psi\varphi$. We concentrate on the symmetric phase for which it is enough to consider the 2PI effective action as a functional of the full propagators $S$ and $D$:
\begin{equation}\label{eq:2PIEA}
\Gamma_{\mbox{\tiny 2PI}}[S,D]=-\sumint\,\Big[\ln S^{-1}-\Sigma S\Big]+\frac{1}{2}\sumint\,\Big[\ln D^{-1}-\Pi D\Big]+\Phi[S,D]\,,
\end{equation}
where $\Phi$ is the sum of 0-leg skeletons (2PI diagrams). When evaluated at its stationary point, the 2PI effective action provides a powerful way to compute the pressure of the system. Further thermodynamic quantities are obtained by taking derivatives. The propagators are given by the stationary condition on $\Gamma_{\mbox{\tiny 2PI}}[S,D]$ which is suitably translated into a set of equations of motion for the self-energies $\Sigma$ and $\Pi$. As an illustration we consider here the one-loop approximation:
\begin{eqnarray}\label{eq:motion_ct}
\Sigma(P) & = & -g^2\sumint_Q D(Q)S(Q+P)-\slash\hspace{-2.5mm}P\delta Z_{\psi}\,,\nonumber\\
\Pi(K) & = & \frac{1}{2}\left(\lambda+\delta\lambda\right)\sumint_Q D(Q)+g^2\mbox{tr}\,\sumint_Q S(Q)S(Q+K)+K^2\delta Z_{\varphi}\,.
\end{eqnarray}
The sum-integrals in the above equations are UV divergent. Renormalization consists in building temperature independent counterterms $\delta Z_\varphi$, $\delta Z_\psi$ and $\delta \lambda$ which absorb all the divergences. In dimensional regularization there is no need for mass counterterms. At one-loop level, the Yukawa coupling is not renormalized and we thus take $\delta g=0$. A general discussion to higher loop orders may be found in \cite{Fermions}.

\section{Renormalization}
One starts focusing on the zero temperature solutions to the equations of motion that we denote by $S_{\rm T=0}$ and $D_{\rm T=0}$. The related self-energies are denoted by $\Sigma_{\rm T=0}$ and $\Pi_{\rm T=0}$ and determine the field strength counterterms $\delta Z_\psi$ and $\delta Z_{\varphi}$ through the renormalization conditions $d\Sigma_{\rm T=0}/d\slash\hspace{-2.5mm}P|_{P_*^2=-\mu^2}=0$ and $d\Pi_{\rm T=0}/dK^2|_{K_*^2=-\mu^2}=0$.

The next step consists in removing coupling singularities from the temperature dependent contributions $\Sigma_{\rm T}=\Sigma-\Sigma_{\rm T=0}$ and $\Pi_{\rm T}=\Pi-\Pi_{\rm T=0}$. It is easy to check that the equation for $\Sigma_{\rm T}$ does not contain any UV divergence. In contrast the equation for $\Pi_{\rm T}$ contains logarithmic divergences. A diagrammatic analysis \cite{Fermions} reveals that these are exactly the same than those encoded in the four-point function with four scalar legs ($\Gamma_{\varphi\varphi}$). Renormalization of these divergences thus amounts to imposing the renormalization condition $\Gamma_{\varphi\varphi}(P_*,K_*)=\lambda$. To proceed, we thus need to know how to build $\Gamma_{\varphi\varphi}$, show that it can be renormalized and check that its renormalization removes the coupling divergences in $\Pi_{\rm T}$.

The function $\Gamma_{\varphi\varphi}$ is built in two steps from a set of nested Bethe-Salpeter equations. These equations involve 2PI {\it kernels} which are directly related to derivatives of the functional $\Phi$ namely $\Lambda_{\psi\psi}=-\delta^2 \Phi/\delta S\delta S$, $\Lambda_{\psi\varphi}=-2\delta^2 \Phi/\delta S\delta D$,  $\Lambda_{\varphi\psi}=-2\delta^2 \Phi/\delta D\delta S^{\rm t}$ and $\Lambda_{\varphi\varphi}=4\delta^2 \Phi/\delta D\delta D$. In the first step, one builds up the four-point function with four fermionic legs $\Gamma_{\psi\psi}$ from $\Lambda_{\psi\psi}$:
\begin{equation}\label{eq:BS1}
\Gamma_{\psi\psi}(P,K)=\Lambda_{\psi\psi}(P,K)-\int_Q\, \Lambda_{\psi\psi}(P,Q)M(Q)\Gamma_{\psi\psi}(Q,K)\,.
\end{equation}
$\Gamma_{\psi\psi}$ is then combined with the 2PI kernels in order to generate a new kernel:
 \begin{eqnarray}\label{eq:link}
\tilde\Lambda_{\varphi\varphi}(P,K) & = & \Lambda_{\varphi\varphi}(P,K)+\int_Q\,\Lambda_{\varphi\psi}(P,Q)M(Q)\Lambda_{\psi\varphi}(Q,K)\nonumber\\
& - & \int_Q\int_R \,\Lambda_{\varphi\psi}(P,Q)M(Q)\Gamma_{\psi\psi}(Q,R)M(R)\Lambda_{\psi\varphi}(R,K)\,.
\end{eqnarray}
The explicit distribution of fermionic indices in Eqs. (\ref{eq:BS1}) and (\ref{eq:link}) can be found in \cite{Fermions} where we also define $M$ as $M_{(\alpha\beta),(\gamma\delta)}(Q)=S_{\alpha\gamma}(Q)S_{\delta\beta}(Q)$. In the second step, a second Bethe-Salpeter equation builds up $\Gamma_{\varphi\varphi}$ from $\tilde\Lambda_{\varphi\varphi}$:
\begin{equation}\label{eq:BS2}
\Gamma_{\varphi\varphi}(P,K)=\tilde\Lambda_{\varphi\varphi}(P,K)-\int_Q\, \tilde\Lambda_{\varphi\varphi}(P,Q)D^2(Q)\Gamma_{\varphi\varphi}(Q,K)\,.
\end{equation} 
At one loop level, it is simple to check that equation (\ref{eq:BS1}) is finite. In contrast, Eq. (\ref{eq:link}) contains logarithmic divergences. However these are overall divergences from which it follows that differences such that $\tilde\Lambda_{\varphi\varphi}(P,Q)-\tilde\Lambda_{\varphi\varphi}(K,Q)$ are finite. Furthermore, by construction, the kernel $\tilde\Lambda_{\varphi\varphi}$ is 2PI with respect to scalar lines implying that $\tilde\Lambda_{\varphi\varphi}(P,Q)-\tilde\Lambda_{\varphi\varphi}(K,Q)\sim 1/Q$ at large $Q$ and fixed $P$ and $K$ \cite{Fermions}. It is then possible to rewrite the equation for $\Gamma_{\varphi\varphi}$ in a UV finite form:
\begin{eqnarray}\label{eq:BS_ren}
\Gamma_{\varphi\varphi}(P,K)-\Gamma_{\varphi\varphi}(P_*,K_*) & = & \tilde\Lambda_{\varphi\varphi}(P,K)-\tilde\Lambda_{\varphi\varphi}(P_*,K_*)\\
& - & \frac{1}{2}\int_Q\Big\{\tilde\Lambda_{\varphi\varphi}(P,Q)-\tilde\Lambda_{\varphi\varphi}(P_*,Q)\Big\}D^2(Q)\Gamma_{\varphi\varphi}(Q,K)\nonumber\\
& - & \frac{1}{2}\int_Q\Gamma_{\varphi\varphi}(P_*,Q)D^2(Q)\Big\{\tilde\Lambda_{\varphi\varphi}(Q,K)-\tilde\Lambda_{\varphi\varphi}(Q,K_*)\Big\}\,.\nonumber
\end{eqnarray}
Using that $\Gamma_{\varphi\varphi}(Q,K)\sim\log Q$, $D(Q)\sim 1/Q^2$ and $\tilde\Lambda_{\varphi\varphi}(Q,K)-\tilde\Lambda_{\varphi\varphi}(Q,K_*)\sim 1/Q$ at large $Q$, one checks that the integrals in (\ref{eq:BS_ren}) are UV finite. Thus $\Gamma_{\varphi\varphi}$ is finite, as announced.

Finally, in order to show that renormalization of $\Gamma_{\varphi\varphi}$ simultaneously removes the coupling divergences in the equation for $\Pi_{\rm T}$, it is convenient to express the latter in terms of $\tilde\Lambda_{\varphi\varphi}$. Followings the steps presented in \cite{Fermions}, one obtains:
\begin{eqnarray}\label{eq:gap_closed}
\Pi_{\rm T}(K) & = & \frac{1}{2}\int_Q\tilde\Lambda_{\varphi\varphi}(K,Q)\delta D(Q)+\frac{1}{2}\int_{\tilde Q}\tilde\Lambda_{\varphi\varphi}(K,{\tilde Q})\sigma_\varphi({\tilde Q})+\Pi_r(K)\,,
\end{eqnarray}
where we have performed the Matsubara sum in order to separate implicit and explicit thermal dependences. The first integral in (\ref{eq:gap_closed}) involves $\delta D(Q)=D(Q)-D_{\rm T=0}(Q)=-\Pi_{\rm T}(Q)D^2(Q)+D_r(Q)$ with $D_r(Q)\sim 1/Q^6$ at large $Q$. The second integral involves $\sigma_\varphi({\tilde Q})=\epsilon(q_0)n(|q_0|)\rho_{\varphi}(q_0,q)$, a particular combination of the sign function $\epsilon(q_0)$, the scalar thermal factor $n(|q_0|)$ and the scalar spectral density $\rho_{\varphi}(q_0,q)$. $\tilde Q$ designates integration along the real axis in contrast to $Q$ which designates integration along the imaginary axis. Finally $\Pi_r(K)$ is a finite function which decreases as $\sim 1/K^2$ at large $K$. Using Eqs. (\ref{eq:BS2}) and (\ref{eq:gap_closed}) one shows that
\begin{equation}\label{eq:ren}
\Pi_{\rm T}(K)=\Pi_r(K)+\frac{1}{2}\int_{\tilde Q}\Gamma_{\varphi\varphi}(K,\tilde Q)\sigma_{\varphi}(\tilde Q)+\frac{1}{2}\int_Q\Gamma_{\varphi\varphi}(K,Q)\Big\{D_r(Q)-D^2(Q)\Pi_r(Q)\Big\}\,.
\end{equation}
Using the finiteness of $\Gamma_{\varphi\varphi}$ and the asymptotic properties of $\Gamma_{\varphi\varphi}$, $\sigma_{\varphi}$, $D_r$ and $\Pi_r$, it is simple to check that this last equation is finite, as expected.

\section{Conclusions}\label{concl}
We have shown on a particular example how to implement renormalization of $\Phi$-derivable approximations in theories with multiple fields (see also \cite{Fermions}). This work is particularly important for gauge theories where one needs to disentangle the UV divergences related to gauge, matter and ghost fields.

\section*{Acknowledgments}
I would like to thank Jean-Paul Blaizot and Julien Serreau for fruitful discussions on this topic and Anton Rebhan for enabling the presentation of this work at the {\it Workshop on Quark-Gluon Plasma Thermalization} (TU Vienna, August 10-12, 2005).

\vfill\eject
\end{document}